\documentclass[aps,pra,twocolumn,superscriptaddress,showpacs,showkeys,amsmath,amssymb]{revtex4}

\usepackage{amsfonts}
\usepackage{amssymb,amsmath}
\usepackage{mathrsfs}
\usepackage{latexsym}
\usepackage{amsmath}
\usepackage[cp1251]{inputenc}
\usepackage{graphicx}
\usepackage{dcolumn}
\usepackage{bm}
\usepackage{color}

\RequirePackage{ifthen}
\RequirePackage[pdfstartview=FitH]{hyperref}
\begin{document}
	
	\title{`Triviality' of universal relations for disordered systems}
   	\author{V.~Pastukhov\footnote{e-mail: volodyapastukhov@gmail.com}}
	\affiliation{Professor Ivan Vakarchuk Department for Theoretical Physics, Ivan Franko National University of Lviv, 12 Drahomanov Street, Lviv, Ukraine}

	\date{\today}

	
	
	\begin{abstract}
	The universal relations for spin-$1/2$ fermions with contact interaction in the presence of quenched disorder are discussed. The disorder is modeled by a random external potential with the Gaussian distribution and $\delta$-like two-point correlation function. Utilizing simple scaling arguments, the renormalizability of the theory, and the Hellmann-Feynman theorem we identified the large-momentum tail of particle distribution and analog of Tan's energy relation for many-fermion systems with disorder in arbitrary dimension $d\ge 2$. It is shown that the energy-pressure relation manifests a kind of scale anomaly in two and three dimensions.
	\end{abstract}
	
	\maketitle
\section{Introduction}
Almost two decades ago Tan published a preprint, which later became a paper \cite{Tan_1}, where a now well-known relation for the energy of spin-$1/2$ fermions interacting through contact two-body potential was proposed. Particularly, he demonstrated that the internal energy of an interacting fermionic system in any thermodynamic phase is a linear functional of the single-particle momentum distribution. Moreover, its large-momentum tail is universal \cite{Tan_2}, $1/|{\bf k}|^4$, with the constant prefactor, called contact, defined as a derivative of energy with respect to the $s$-wave scattering length. Later on, Braaten and Platter \cite{Braaten_2008} rederived these relations using a technique from the quantum field theory known as the operator product expansion. At that time, the large-momentum tail for the one-dimensional exactly solvable Lieb-Liniger model was known \cite{Olshanii_2003}. Further developments (see \cite{Strinati_2018} for review) extended these results to one- \cite{Barth_2011,Patu_2017}, two- \cite{Valiente_2011,Werner_2012} and, in general, arbitrary $d$-dimensional \cite{Valiente_2012} geometries. A field-theoretical derivation of Tan`s relations for mixtures of atoms with arbitrary mass ratios and compositions of constituents was provided in Ref.~\cite{Combescot_2009}. The universality of ultracold atomic gases with short-range potentials is not only a theoretical concept. The experimental verification of Tan's relations came shortly \cite{Stewart_2010,Kuhnle_2010} after the theory development. Moreover, most of the thermodynamic properties of systems with zero-range interaction can be obtained through the precise measurements \cite{Carcy_2019,Mukherjee_2019,Jager_2024} or utilizing theoretical calculation methods, including $t$-matrix approximation \cite{Palestini_2010}, high-temperature expansion \cite{Sun_2015,Hou_2020}, functional renormalization group \cite{Boettcher_2013}, Monte Carlo simulations \cite{Drut_2010,Rossi_2018,Jensen_2020} and lattice formulation \cite{Alexandru_2021}, of the Tan contact. This parameter also governs \cite{Hu_2009,Hoinka_2013} the ultraviolet behavior of the static structure factor, and determines fraction \cite{Werner_2009,Zhang_2009} of the closed-channel molecules in a system of spin-$1/2$ fermions near the Feshbach resonance. Furthermore, Tan's contact controls hydrodynamic characteristics \cite{Hofmann_2011} and the large-momentum behavior of the quasiparticle excitation spectrum and its damping rate \cite{Nishida_2012} in the many-body limit. The concept of universality is not restricted to contact interaction. Recently, universal relations were discussed in the context of dipolar gases \cite{Hofmann_2021,Cherny_2023}, and dilute systems with the power-law two-body potentials \cite{Voronova_2024}.

In this paper, we obtain the universal relations for disordered spinfull fermions with contact two-body interaction in arbitrary dimensions. For this model, the disorder impact increases when the system undergoes from the Bardeen-Cooper-Schrieffer to Bose-Einstein condensation sides of crossover both in superfluid \cite{Orso_2007,Han_2011,Khan_2012} and normal \cite{Palestini_2013} phases in three and two dimensions \cite{Khan_2017}. Although the quenched disorder is modeled by random external potential, its impact on properties of the system does not trivially reduce to generalized virial theorem \cite{Tan_3}. A physical idea underlying the quench disorder is two subsystems with totally different (in magnitude) relaxation times. Alternatively, the disorder can be thought as an additional effective thermostat \cite{Haga_2020} breaking quantum coherence even at absolute zero. Physically the most relevant model for the disorder is the infinitely heavy (classical) non-interacting particles (impurities) immersed in quantum system. The integration over the coordinates of the classical subsystem provides the disorder averaging. If one assumes the contact interaction between particles of quantum subsystem and impurities, the universal relations is straightforwardly obtained considering a mixture of quantum subsystem and bosons \cite{Werner_2012_2} and then taking the limit of infinite boson mass. The discussed below case of the Gaussian $\delta$-correlated disorder, although conceptually simpler, provides more tricky results.

\section{Formulation}
\subsection{Model description}
For concreteness, we consider $N$ spin-$1/2$ fermions with contact two-body interaction. The results, however, can be straightforwardly applied to bosons. The $d$-dimensional system is loaded in a large volume $L^d$ with the periodic boundary conditions and the random external potential $V({\bf r})$ acting on every atom. This model is described by the following Euclidean action:
\begin{eqnarray}\label{S}
S=\int_x\psi^{\dagger}_{\sigma}\left\{\partial_{\tau}-\xi
-V\right\}\psi_{\sigma}-g_{\Lambda}\int_x\psi^{\dagger}_{\uparrow}\psi^{\dagger}_{\downarrow}\psi_{\downarrow}\psi_{\uparrow},
\end{eqnarray}
where the Grassmann fields $\psi^{\dagger}_{\sigma}(x)$, $\psi_{\sigma}(x)$ describe fermions in spin state $\sigma$ (from now on we will adopt the summation convention) and with quadratic dispersion $\xi=\varepsilon-\mu$ (here $\varepsilon=-\frac{\nabla^2}{2m}$ and $\mu$ is the chemical potential that fixes the total density) and $\int_x=\int^{\beta}_0 d\tau\int_{L^d} d{\bf r}$ (where $\beta$ is the inverse temperature) denotes the integration in the $(d+1)$-dimensional Euclidean space-time. Note that the populations of fermions in different spin states are not important for the consideration below. Moreover, the phase of interacting fermions, being it the Fermi liquid or superfluid states, is also irrelevant. A bare coupling constant $g_{\Lambda}$ depends on the ultraviolet (UV) cutoff parameter $\Lambda$ and is related to the `observable' coupling $g$ through the equation (see, for instance, \cite{Hryhorchak_2023})
\begin{eqnarray}\label{g_Lambda}
\frac{1}{g_{\Lambda}}=\frac{1}{g}-\frac{1}{L^d}\sum_{|{\bf k}|\le \Lambda}\frac{1}{2\varepsilon_k}.
\end{eqnarray}
The latter parameter $g$ controls the strength of the $s$-wave two-body interaction among fermions in the different spin states. For $g>0$ there is a single bound state with the energy, $\epsilon_a=-\frac{1}{ma^2}$, related 
\begin{eqnarray}\label{g}
g^{-1}=-\frac{\Gamma(1-d/2)}{(4\pi)^{d/2}}m^{d/2}|\epsilon_a|^{d/2-1},
\end{eqnarray}
to the `observable' coupling. In 2$d$, the sum over the wave vector in Eq.~(\ref{g_Lambda}) is not only UV, but also infrared divergent. The bare coupling, therefore, is determined as follows $g^{-1}_{\Lambda}=-\frac{1}{L^2}\sum_{|{\bf k}|\le \Lambda}\frac{1}{2\varepsilon_k+|\epsilon_a|}$ in two dimensions. In principle, such a form is valid in any dimension for positive $g$s. For simplicity, the distribution functional, $P[V]\propto \exp\left\{-\frac{1}{2h^2}\int_{\bf r}V^2({\bf r})\right\}$, describing fluctuations of $V({\bf r})$ is assumed to be Gaussian and $\delta$-correlated, with the averages defined as follows
\begin{align}\label{over_V}
\overline{V({\bf r})}=0, \ \ \overline{V({\bf r})V({\bf r}')}=h^2\delta({\bf r}-{\bf r}').
\end{align}
The disorder strength is characterized by a single constant $h$ (with dimension of $\frac{\textrm{length}^{d/2-2}}{\textrm{mass}}$). Implicitly, the presence of the UV cutoff impacts the disorder definition. Particularly, the theory's locality is smeared below length scale $\Lambda^{-1}$, such that there are no infinities in correlators formed by the products of an arbitrary number of fields at short distances. This means that all sums over the wave vectors are restricted (although not explicitly written below) by the upper summation limit $\Lambda$, in particular, the $\delta$-function in (\ref{over_V}) should be treated, $\delta_{\Lambda}({\bf r})=\frac{1}{L^d}\sum_{|{\bf k}|\le \Lambda}e^{i{\bf k}{\bf r}}$, as a finite function at the origin.

\subsection{Single particle in random potential}
\begin{figure}[h!]
	\centerline{\includegraphics
		[width=0.4750
		\textwidth,clip,angle=-0]{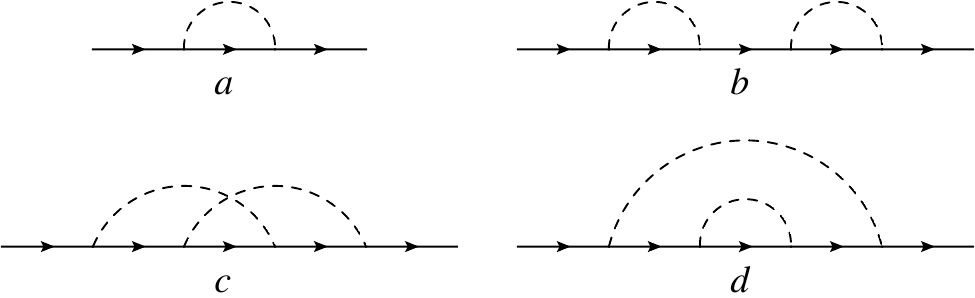}}
	\caption{First- $(a)$ and second-order $(b-d)$ corrections to the disorder-averaged Green's function.}\label{disorder_12_fig}
\end{figure}

Before we move on to the thermodynamic limit, let us first overview (see \cite{Mahan_2000}) the one-body problem in the random potential. In this case $\mu=0$, and the interaction term in action $S$ is turned off. The one-fermion Green's function $-\langle\psi(x)\psi^{\dagger}(x')\rangle=\frac{1}{\beta}\sum_{\nu_n}e^{i\nu_n\tau}\langle {\bf r}|G(i\nu_n)|{\bf r}'\rangle$ (the spin index is temporarily eliminated), written down via a sum over the fermionic Matsubara frequency in momentum basis, reads
\begin{align}\label{G}
\langle {\bf k}|G(i\nu_n)|{\bf k}'\rangle=\langle {\bf k}|\frac{1}{i\nu_n-\varepsilon-V}|{\bf k}'\rangle.
\end{align}
The disorder averaging is usually performed by expanding the denominator of (\ref{G}) in powers of $V$. Leaving only non-zero (with an even number of $V$s) terms
\begin{align}\label{over_G}
&\overline{G(i\nu_n)}=G_0(i\nu_n)+G_0(i\nu_n)\overline{VG_0(i\nu_n)V}G_0(i\nu_n)\nonumber\\
&+G_0(i\nu_n)\overline{VG_0(i\nu_n)VG_0(i\nu_n)VG_0(i\nu_n)V}G_0(i\nu_n)+\dots,
\end{align}
where $G_0(i\nu_n)=\frac{1}{i\nu_n-\varepsilon}$. Making use of the closure relation $1=\sum_{\bf k}|{\bf k}\rangle\langle{\bf k}|$ by putting it in between every pair of standing next to each other $V$ and $G_0(i\nu_n)$ operators, the disorder averaging reduces to a kind of Wick's theorem contracting two matrix elements of potential $V$
\begin{align}\label{V_Wick}
\overline{\langle{\bf k}|V|{\bf q}\rangle\langle{\bf k}'|V|{\bf q}'\rangle}=\frac{h^2}{L^d}\delta_{{\bf k}+{\bf k}',{\bf q}+{\bf q}'},
\end{align}
(where $\delta_{{\bf k},{\bf q}}$ is a product of $d$ Kronecker deltas)  in accordance with (\ref{over_V}). This remark allows us to visualize the disorder-averaging procedure by introducing graphical representation \cite{Edwards_1958}. If we associate straight line with arrow carrying momentum ${\bf k}$ with the diagonal matrix element of $\langle{\bf k}|G_0(i\nu_n)|{\bf k}\rangle=\frac{1}{i\nu_n-\varepsilon_{\bf k}}$ and dashed one with the contraction (\ref{V_Wick}). Then, the first correction to $\overline{\langle {\bf k}|G(i\nu_n)|{\bf k}\rangle}$ (note that only diagonal elements survive after restoring of the translation invariance by the disorder averaging) can be represented as shown in Fig.~\ref{disorder_12_fig} $(a)$.
The disorder averaging of the second correction to $\langle{\bf k}|G(i\nu_n)|{\bf k}\rangle$ gives three different graphs $(b)-(d)$ in Fig.~\ref{disorder_12_fig}.
Moving further in perturbation theory it is clear that the number of terms increases significantly in every next order. However, all graphs can be divided into two classes: the first one is the single-particle reducible, i.e. can be split into two by cutting a single horizontal line (in Fig.~\ref{disorder_12_fig} there is only one $(a)$ such diagram); while the second class contains all other (one-particle-irreducible) graphs. By collecting all reducible diagrams in every order of perturbation theory, and adding the result to the zero-order (free-particle) Green's function, one obtains 
\begin{align}\label{}
\overline{\langle {\bf k}|G(i\nu_n)|{\bf k}\rangle}=G_{\bf k}(i\nu_n)=\frac{1}{i\nu_n-\varepsilon_{\bf k}-\Sigma_{\bf k}(i\nu_n)},
\end{align}
with the self-energy $\Sigma_{\bf k}(i\nu_n)$ given by a sum of all one-particle irreducible graphs (without inclusion of external lines). The further reduction of the graphs' number can be attained by the line thickening, i.e. the summation of the self-energy insertions in the internal Green's functions. The simplest example of such an insertion is presented by diagram $(d)$ in Fig.~\ref{disorder_12_fig}. In higher orders of perturbation theory, there are terms with the insertions of all the other graphs forming $\Sigma_{\bf k}(i\nu_n)$. This procedure known as a skeleton-graph summation, further reduces the number of corrections to the self-energy
\begin{align}\label{Sigma_skeleton}
&\Sigma_{\bf k}(i\nu_n)=\frac{h^2}{L^d}\sum_{\bf q}G_{\bf q}(i\nu_n)\nonumber\\
&+\frac{h^4}{L^{2d}}\sum_{{\bf q},{\bf q}'}G_{\bf q}(i\nu_n)G_{{\bf q}'}(i\nu_n)G_{{\bf q}+{\bf q}'-{\bf k}}(i\nu_n)+\dots.
\end{align}
Note that here we must throw away from the initial self-energy (with zero-order lines $\langle{\bf k}|G_0(i\nu_n)|{\bf k}\rangle$) all the graphs with the self-energy insertions in internal lines, which sometimes is a complicated task in higher orders of perturbation theory. Together with $G_{\bf k}(i\nu_n)$ given by the Dyson equation, Eq.~(\ref{Sigma_skeleton}) represents a complicated nonlinear integral equation with a non-perturbative nature. Since dots in (\ref{Sigma_skeleton}) contain an infinite number of terms, it is impossible to solve this equation exactly. However, the leading-order UV behavior of the disordered-averaged Green function is accessible. The latter is important for the renormalization procedure. Indeed, the first and the second terms in (\ref{Sigma_skeleton}) diverge in the limit $\Lambda\to \infty$ for $d\ge 2$ and $d\ge 3$, respectively. For higher (unphysical) dimensions, the number of UV-divergent terms grows such that exactly in $d=4$ the whole series is blowing up.
A power counting fixes the leading-order $\Lambda$-dependence of the first two terms in (\ref{Sigma_skeleton}): $\Lambda^{d-2}$ and $\Lambda^{2(d-3)}$, respectively. It is possible, however, to obtain the appropriate estimation for every term is series for $\Sigma_{\bf k}(i\nu_n)$. It is obvious that for a graph with $l$ dashed lines, there is $2l-1$ internal $G$-lines, and consequently $2l-1$ momentum summations. From the latter number, we must subtract $l-1$ independent $\delta$-symbols, which leaves us with $l$ different wave-vector summations. Then, the power of divergence for the considered diagram is $ld-2(2l-1)=2-(4-d)l$, and if this number $2-(4-d)l\ge 0$, the diagram diverges. It should be noted that this estimation is valid only for the skeleton graphs (i.e. without the self-energy insertions). The obtained inequality says that in $2d$ we have only one (with $l=1$) logarithmically-divergent diagram, while in $3d$ two terms are divergent: the first one linearly and the second one logarithmically in $\Lambda$. To cure infinities one should adopt some renormalization procedure. The simplest one is dimensional regularization, but one cannot eliminate large logarithms within this prescription. Therefore here we use the minimal subtraction scheme by introducing counterterms. The most natural way to introduce them is to redefine random potential $V({\bf r})\to\overline{V}_\Lambda+\delta V({\bf r})$, where $\overline{V({\bf r})}=\overline{V}_\Lambda$ absorbs all divergences and $\overline{\delta V({\bf r})\delta V({\bf r}')}=h^2\delta({\bf r}-{\bf r}')$. It is straightforward to argue that the leading-order large-$k$ behavior, $k^{2(d-3)}$, of the regularized self-energy is dictated by the second term in Eq.~(\ref{Sigma_skeleton}). In all dimensions $d<4$ this asymptotics is subleading in comparison to free particle dispersion, which is an important fact for the correct regularization. The requirement of independence of Green's function on the UV cutoff $\partial_{\Lambda} G_{\bf k}(i\nu_n)=0$ [or, equivalently  $\partial_{\Lambda} \overline{V}_\Lambda =-\partial_{\Lambda} \Sigma_{\bf k}(i\nu_n)$] fixes, up to a constant, the counterterm $\overline{V}_\Lambda$
\begin{align}
&\Lambda\partial_{\Lambda} \overline{V}_\Lambda=\frac{\Lambda^2}{2m}c^{(1)}_d\left[(2mh)^2\Lambda^{d-4}+\dots\right], \ \  (2\le d<3), \label{counterm_2}\\
&\Lambda\partial_{\Lambda} \overline{V}_\Lambda=\frac{\Lambda^2}{2m}c^{(1)}_d\left[(2mh)^2\Lambda^{d-4}\right.\nonumber\\
&\left.+c^{(2)}_d(2mh)^4\Lambda^{2(d-4)}+\dots\right], \ \  (3\le d<3\frac{1}{3}),\label{counterm_3}
\end{align}
where numerical coefficients $c^{(1)}_d=\frac{2^{1-d}}{\pi^{d/2}\Gamma(d/2)}$ [area of unit sphere in $d$ dimensions divided by $(2\pi)^d$] and  $1/c^{(2)}_d=2^d(4\pi)^{\frac{d-3}{2}}\Gamma\left(\frac{d-1}{2}\right)\sin(\pi[d/2-1])$, and $\dots$ denotes vanishing in the limit $\Lambda\to \infty$ terms. In higher dimensions, the number of terms in $\overline{V}_\Lambda$ increases becoming infinite exactly in 4$d$. For the integration of Eqs.~(\ref{counterm_2}),(\ref{counterm_3}) in physical dimensions $d=2$ and $d=3$, one must explicitly introduce the characteristic length scale $b$ of the disorder potential such that $h\propto\frac{b^{d/2-2}}{m}$. Then, a simple integration leaves us with the result
\begin{align}
&\overline{V}_\Lambda=\frac{mh^2}{\pi}\ln(\Lambda b), \ \  (d=2),\\
&\overline{V}_\Lambda=\frac{mh^2}{\pi^2}\Lambda+\frac{m^3h^4}{2\pi^2}\ln(\Lambda b), \ \ (d=3).
\end{align}
These counterterms are determined up to constant $\propto\frac{1}{mb^2}$ constructed from dimensional arguments. However, adding them leaves the further discussion unaffected.

\section{Universal relations}
\subsection{Contacts}
A mathematical program of the quenched disorder realization is the following: first, we have to compute the partition function $Z=\int D\psi^{\dagger}_{\sigma} D\psi_{\sigma}e^S$ of the system (fermions in our case) in the presence of random potential $V({\bf r})$, and then identify the thermodynamic potential with averaging of its logarithm, i.e. $-\beta\Omega=\overline{\ln Z}$. The same applies to various correlation functions, i.e. the recipe reads: $\overline{\langle\dots\rangle}$. The first step of these calculations is complicated by the broken continuous translation invariance, which however is restored in the second one. This is not very convenient though, and to overcome the symmetry breaking one typically utilizes the replica trick \cite{Edwards_1971}. An idea behind replicas is the formal equality $\overline{\ln Z}=\left.\partial_{\mathcal{N}}\overline{Z^\mathcal{N}}\right|_{\mathcal{N}=0}$, prompting us to consider $\mathcal{N}$ identical copies of the system described by action (\ref{S}) each, with the subsequent taking of limit $\mathcal{N}\to 0$ at the end of calculations. More details in the context of the considered system can be found in Ref.~\cite{Han_2011}. This prescription allows one to use the Hellmann–Feynman relations (theorems). Keeping in mind that with the counterterms $\overline{V}_\Lambda$ and bare coupling $g_{\Lambda}$ being introduced, the many-body theory is well defined, we are free to differentiate the thermodynamic potential with respect to `observable' coupling constants $g$, $h$ and mass of particles $m$ expecting finite results. In particular, from the mass derivative which determines the average (both statistical and over the disorder) kinetic energy of the system
\begin{align}
m\partial_m\left(-\beta\Omega\right)=m\partial_m\overline{\ln Z}=m\partial_m\left.\partial_{\mathcal{N}}\overline{Z^\mathcal{N}}\right|_{\mathcal{N}=0},
\end{align}
(where both $h$ and $g$ are kept fixed) we can obtain the following identity:
\begin{align}
m\partial_m\left(-\frac{\Omega}{L^d}\right)=\overline{\langle\psi^{\dagger}_{\sigma}\varepsilon \psi_{\sigma}\rangle}-\overline{\langle\psi^{\dagger}_{\sigma}\psi_{\sigma}\rangle}m\partial_m\overline{V}_{\Lambda}\nonumber\\
-\overline{\langle\psi^{\dagger}_{\uparrow}\psi^{\dagger}_{\downarrow}\psi_{\downarrow}\psi_{\uparrow}\rangle}m\partial_mg_{\Lambda}.
\end{align}
Recall the r.h.s. of the above expression is finite in the $\Lambda\to \infty$ limit. This fixes the large-momentum behavior of the particle distribution. Making use of derivative $m\partial_mg_{\Lambda}=g^2_{\Lambda}\frac{1}{L^d}\sum_{{\bf k}}\frac{1}{2\varepsilon_k}$ and differentiating counterterms (\ref{counterm_2}), (\ref{counterm_3}), we obtain for the balanced mixture ($\overline{\langle\psi^{\dagger}_{\uparrow,\downarrow}\psi_{\uparrow,\downarrow}\rangle}=n$)
\begin{eqnarray}\label{N_k_sigma}
\overline{N}_{{\bf k}}=\frac{\overline{\mathcal{C}}_g+\mathcal{C}^{(1)}_h}{|{\bf k}|^4}+\frac{\mathcal{C}^{(2)}_h}{|{\bf k}|^{8-d}}+\dots, \ \ (|{\bf k}|\to \infty),
\end{eqnarray}
where $\overline{\mathcal{C}}_g=(mg_{\Lambda})^2\overline{\langle\psi^{\dagger}_{\uparrow}\psi^{\dagger}_{\downarrow}\psi_{\downarrow}\psi_{\uparrow}\rangle}$ is a standard disorder-averaged Tan's contact while two other parameters
\begin{eqnarray}\label{C_h}
\mathcal{C}^{(1)}_h=(2mh)^2n, \ \  \mathcal{C}^{(2)}_h=3c^{(2)}_d(2mh)^4n,
\end{eqnarray} 
linearly depend on density. For imbalanced case, contacts $\mathcal{C}^{(1)}_h$, $\mathcal{C}^{(2)}_h$ should be introduced for every spin component and are defined by very similar formulas with the replacement $n\to n_{\uparrow,\downarrow}$. Next, taking the derivative of the $\Omega$-potential with respect to the two-body coupling
\begin{align}\label{C_g}
g\partial_g\frac{\Omega}{L^d}=\overline{\langle\psi^{\dagger}_{\uparrow}\psi^{\dagger}_{\downarrow}\psi_{\downarrow}\psi_{\uparrow}\rangle}g\partial_gg_{\Lambda}=\frac{\overline{\mathcal{C}}_g}{m^2g},
\end{align}
one rederives the well-known \cite{Tan_2} adiabatic sweep theorem, trivially generalized here to disordered systems. Explicitly calculating derivative $g\partial_gg_{\Lambda}=\frac{g^2_{\Lambda}}{g}$ we again convince ourselves that the product $g^2_{\Lambda}\overline{\langle\psi^{\dagger}_{\uparrow}\psi^{\dagger}_{\downarrow}\psi_{\downarrow}\psi_{\uparrow}\rangle}$ remains finite at infinite UV cutoff. The calculations of the derivative of the thermodynamic potential with respect to coupling $h$, determining the strength of the disorder, are more tricky. In this case, it is convenient to introduce a new variable $v({\bf r})=\delta V({\bf r})/h$ in the functional integral defining the disorder averaging, and then, take the $\partial_h$-derivative. After this, the computations are straightforward and give
\begin{align}\label{dh_Omega}
h\partial_h\frac{\Omega}{L^d}=\overline{\langle\psi^{\dagger}_{\sigma}\psi_{\sigma}\rangle}h\partial_h\overline{V}_{\Lambda}+\overline{\delta V\langle\psi^{\dagger}_{\sigma}\psi_{\sigma}\rangle}.
\end{align}
Importantly that a local correlator, $\overline{\delta V\langle\psi^{\dagger}_{\sigma}\psi_{\sigma}\rangle}$, is not well-defined at large UV cutoffs. Of course, there are equivalent forms, involving somewhat different correlators, for writing down the last equality. From a physical point of view, correlator $\overline{\delta V\langle\psi^{\dagger}_{\sigma}\psi_{\sigma}\rangle}$ represents the average response of the local density of fermions to random potential. Introducing notation $\rho_{\bf k}=\frac{1}{\sqrt{\beta L^d}}\int_xe^{-i{\bf k}{\bf r}}\psi^{\dagger}_{\sigma}\psi_{\sigma}$ (${\bf k}\neq 0$), one proves, utilizing replicated action, the following identity: $\overline{\delta V\langle\psi^{\dagger}_{\sigma}\psi_{\sigma}\rangle}=-\frac{h^2}{L^d}\sum_{\bf k}\overline{\langle\rho_{\bf k}\rho_{-{\bf k}}\rangle}$. A meaning of $\overline{\langle\rho_{\bf k}\rho_{-{\bf k}}\rangle}$, as it results from the definition, is the zero Matsubara frequency limit of full (recall the sum over $\sigma$ in the formula for $\rho_{\bf k}$) dynamic structure factor of the disordered interacting system. Taking into account the first $\Lambda$-dependent term in Eq.~(\ref{dh_Omega}) and finiteness of $h\partial_h\Omega$ at infinite cutoff, one elementarily concludes
\begin{eqnarray}\label{rho_k}
\overline{\langle\rho_{\bf k}\rho_{-{\bf k}}\rangle}|_{|{\bf k}|\to \infty}=\frac{\mathcal{C}^{(1)}_h/h^2}{m|{\bf k}|^2}+\frac{4\mathcal{C}^{(2)}_h/h^2}{3m|{\bf k}|^{6-d}}+\dots.
\end{eqnarray}
Yet another observable quantity, except $\overline{N}_{{\bf k}}$, demonstrates the universal large-momentum behavior with a linear dependence on the system's density. Note that the first term in (\ref{rho_k}) is standard $2n/\varepsilon_{\bf k}$, coming from general arguments like sum rules (note that $2n$ is the total density of the system), while the second one is disorder-induced.
Combining all the obtained identities, one can rewrite the energy density of the system as follows:
\begin{align}\label{E_density}
&\frac{E}{L^d}=\overline{\langle\psi^{\dagger}_{\sigma}\varepsilon \psi_{\sigma}\rangle}+\overline{V}_\Lambda\overline{\langle\psi^{\dagger}_{\sigma}\psi_{\sigma}\rangle}+\overline{\delta V\langle\psi^{\dagger}_{\sigma}\psi_{\sigma}\rangle}\nonumber\\
&+g_{\Lambda}\overline{\langle\psi^{\dagger}_{\uparrow}\psi^{\dagger}_{\downarrow}\psi_{\downarrow}\psi_{\uparrow}\rangle}
=\left(g\partial_g-m\partial_m+h\partial_h\right)\frac{\Omega}{L^d},
\end{align}
where in the last equality the not written down residual term, $\overline{\langle\psi^{\dagger}_{\sigma}\psi_{\sigma}\rangle}\left(h\partial_h-m\partial_m-1\right)\overline{V}_\Lambda$, equals zero identically. The latter fact can be confirmed by direct calculations using counterterms or a general UV structure of the self-energy (\ref{Sigma_skeleton}) in higher dimensions.

Till now we have been working in the grand-canonical ensemble, i.e. all the derivatives, $\partial_{m,g,h}$, were calculated with the chemical potential $\mu$ kept fixed. However, very often it is more convenient to operate in a canonical ensemble with a constant density of the system. In this case, all the above equations remain true with the replacement $\Omega\to E$ (and the Helmholtz free energy $F$ at finite temperatures) and all derivatives taken at fixed $n$. For instance, being applied to Eq.~(\ref{C_g}), this formal rule gives $g\left(\partial_g\frac{E}{L^d}\right)_n=\frac{\overline{\mathcal{C}}_g}{m^2g}$ a standard definition of the Tan contact. A reformulation of (\ref{E_density}) is also obvious $E=\left(g\partial_g-m\partial_m+h\partial_h\right)_nE$ and not unexpected. Indeed, one can always represent the energy per particle in the form $E/N=\frac{n^{2/d}}{m}\mathcal{E}(an^{1/d},bn^{1/d})$, where $\mathcal{E}(an^{1/d},bn^{1/d})$ is an arbitrary function with a non-trivial zero-density limit $\frac{\#}{mb^2}$ (here $\#$ denotes numerical constant). Noticing that $a^{d-2}= mg$ and $b^{d/2-2}= mh$ (equalities up to numerical prefactors), it is readily to show the above identity. One can introduce the `full' derivative $\frac{d}{d m}=d_m$ with respect to mass and with both $a$ and $b$ kept fixed. Then, an equivalent formula for the energy, which is convenient for the calculations in $2d$, reads $E=-(md_m)_{\mu}\Omega=-(md_m)_{n}E$. 

Getting everything together, we obtain the energy in the conventional Tan-like form
\begin{eqnarray}\label{E_2d}
E&=&\sum_{{\bf k},\sigma}\varepsilon_{\bf k}\left[\overline{N}_{{\bf k}}-\frac{\overline{\mathcal{C}}_g+\mathcal{C}^{(1)}_h}{|{\bf k}|^2(|{\bf k}|^2+a^{-2})}\right]\nonumber\\
&-&h^2\sum_{{\bf k}}\left[\overline{\langle\rho_{\bf k}\rho_{-{\bf k}}\rangle}-\frac{\mathcal{C}^{(1)}_h/mh^2}{|{\bf k}|^2+a^{-2}}\right],
\end{eqnarray}
after some rearrangement in $2d$. The three-dimensional counterpart is
\begin{eqnarray}\label{E_3d}
E&=&\frac{L^3\overline{\mathcal{C}}_g}{4\pi ma}+\sum_{{\bf k},\sigma}\varepsilon_{\bf k}\left[\overline{N}_{{\bf k}}-\frac{\overline{\mathcal{C}}_g+\mathcal{C}^{(1)}_h}{|{\bf k}|^4}\right.\nonumber\\
&&\left.-\frac{\mathcal{C}^{(2)}_h}{|{\bf k}|^4(|{\bf k}|+b^{-1})}\right]-h^2\sum_{{\bf k}}\left[\overline{\langle\rho_{\bf k}\rho_{-{\bf k}}\rangle}\right.\nonumber\\
&&\left.-\frac{\mathcal{C}^{(1)}_h/h^2}{m|{\bf k}|^2}-\frac{4\mathcal{C}^{(2)}_h/3mh^2}{|{\bf k}|^2(|{\bf k}|+b^{-1})}\right].
\end{eqnarray}
Both sums over the wave vectors in Eqs.~(\ref{E_2d}), (\ref{E_3d}) converge. Formally, terms with $\mathcal{C}^{(1)}_h$ fully and with $\mathcal{C}^{(2)}_h$ partially compensate each other leaving sums divergent. Without disorder $h=0$, these equations reproduce well-known Tan's energy theorem from literature \cite{Tan_1,Braaten_2008,Valiente_2011,Werner_2012} (note that our definition of the scattering length in $2d$ differs by a factor $2e^{-\gamma}$, with $\gamma= 0.5772...$ being the Euler-Mascheroni constant). It should be noted that the energy theorem in the presence of disorder involves not only the disorder-averaged momentum distribution $\overline{N}_{{\bf k}}$ and contact parameter $\overline{\mathcal{C}}_g$ but also a zero-frequency two-point correlator $\overline{\langle\rho_{\bf k}\rho_{-{\bf k}}\rangle}$ of the particle number fluctuation operator.

\subsection{Energy-pressure relation}
The pressure and energy density of Galilean-invariant $d$-dimensional ideal quantum gases are related $p=\frac{2}{d}\frac{E}{L^d}$. The same relation holds \cite{Zhang_2009} for the unitary Fermi gas, where the $s$-wave scattering length $a$ diverges and all the physics is uniquely controlled by the density of the system. A more general case of finite $a$s was considered in \cite{Tan_3}. Here we aim at the energy-pressure relation for systems with contact two-body interaction in the presence of quenched $\delta$-correlated disorder. For this purpose, we recall a formula, $\Omega=-pL^d$, from thermodynamics and will use simple scaling arguments in combination with the formal identity for derivative $\partial\Omega/\partial L^d$. In action (\ref{S}), the volume `sits' only in the integration limits of $\int_{L^d} d{\bf r}$, therefore, by rescaling coordinates ${\bf r}=L\tilde{\bf r}$, one can calculate the derivative $\partial_L\Omega$. To do this we must find out the scaling properties of fields and couplings in $S$. The requirement for the total number of particles to be fixed leads us to conclusion $\psi_{\sigma}(\psi_{\sigma}^{\dagger})\to L^{-d/2}\tilde{\psi}_{\sigma}(\tilde{\psi}^{\dagger}_{\sigma})$ (where `tilde' on fields denotes dependence on $\tilde{\bf r}$). In order to identify correct scaling transformation for the random potential $\delta V({\bf r})$, we should keep track of the disorder averaging $\overline{\delta V(L\tilde{\bf r})\delta V(L\tilde{\bf r}')}=L^{-d}h^2\delta(\tilde{\bf r}-\tilde{\bf r}')$. From the latter equation we conclude that $\delta V\to L^{-d/2}\delta \tilde{V}$. An equivalent way to show the last transformation is to single out $\delta V({\bf r})=hv({\bf r})$ explicit dependence on parameter $h$, and then, apply the dimensional arguments. Finally, we must take into account the UV cutoff. Its role is to smear out the lower integration limit in $\int_{L^d} d{\bf r}$ at length scales of order $\Lambda^{-1}$. Then, the change of coordinates sets the dimensionless cutoff $\tilde{\Lambda}=\Lambda L$ (or keep in mind the relation $\Lambda\propto 1/L$). At this point, we are in a position to calculate derivative $\partial\Omega/\partial L^d$ utilizing the path-integral in new `tilded' fields, and then, turn back to the original ones by rescaling. The obtained in such a way formula for pressure
\begin{align}\label{}
p=\frac{2}{d}\overline{\langle\psi^{\dagger}_{\sigma}\varepsilon \psi_{\sigma}\rangle}+\frac{1}{d}\overline{\langle\psi^{\dagger}_{\sigma}\psi_{\sigma}\rangle}\Lambda\partial_{\Lambda}\overline{V}_\Lambda+\frac{1}{2}\overline{\delta V\langle\psi^{\dagger}_{\sigma}\psi_{\sigma}\rangle}\nonumber\\
+\left(g_{\Lambda}+\frac{1}{d}\Lambda\partial_{\Lambda}g_{\Lambda}\right)\overline{\langle\psi^{\dagger}_{\uparrow}\psi^{\dagger}_{\downarrow}\psi_{\downarrow}\psi_{\uparrow}\rangle},
\end{align}
can be equivalently rewritten, with help of the previously derived identities, as follows:
\begin{eqnarray}\label{p}
p&=&\frac{1}{d}\left(2+a\partial_a+b\partial_b\right)\frac{E}{L^d}\\
&+&\frac{1}{d}\overline{\langle\psi^{\dagger}_{\sigma}\psi_{\sigma}\rangle}\left(\Lambda\partial_{\Lambda}-2-b\partial_b\right)\overline{V}_\Lambda,\nonumber
\end{eqnarray}
where the second term is zero identically thanks to $\left(\Lambda\partial_{\Lambda}-2-b\partial_b\right)\overline{V}_\Lambda=0$. The latter is readily shown if one takes advantage of an explicit formula for the counterterm. This is another consistency check for the adopted calculation procedure. Equation~(\ref{p}) is expected from the previous discussion. Indeed, the pressure can be computed from energy $p=-\partial E/\partial L^d$ according to thermodynamics. Then, $p=n^2\partial_n(E/N)=n^2\partial_n\frac{n^{2/d}}{m}\mathcal{E}(an^{1/d},bn^{1/d})=\frac{n}{d}\left(2+a\partial_a+b\partial_b\right)\frac{n^{2/d}}{m}\mathcal{E}(an^{1/d},bn^{1/d})$, which proves (\ref{p}). Likewise Eqs.~(\ref{E_2d}), (\ref{E_3d}), this equation can be rewritten via the contact parameter and sum over wave vector involving correlator $\overline{\langle\rho_{\bf k}\rho_{-{\bf k}}\rangle}$. In two- and three-dimensional geometries, where the counterterms possess logarithms, the pressure contains anomalous terms $\frac{mh^2n}{\pi}$ and $\frac{m^3h^4n}{3\pi^2}$, respectively.

\section{Summary and Discussion}
Concluding, we have obtained universal relations for $d$-dimensional spin-$1/2$ fermions with contact two-body interaction in the presence of quenched disorder. More concretely, the disorder is modeled by random external potential with Gaussian distribution and $\delta$-correlated average. Being renormalizable (even super-renormalizable below $d=4$), this theory requires a finite number of counterterms to be well-defined. These counterterms possess a simple structure and predetermine the short-distance behavior of fields' correlators. In particular, focusing mostly on the two- and three-dimensional cases, we have shown that the large momentum behavior of the particle distribution function is controlled, apart from the disorder-average Tan's contact, by a number (depending on spatial dimension) of simple constants that linearly depend on the density of the system.

A better understanding of the obtained large-momentum tail in the particle distribution can be already realized considering spin-polarized ideal Fermi gas in random potential. The momentum distribution in this case is given by a standard formula
\begin{eqnarray}\label{N_k}
\overline{N}_{\bf k}=\frac{1}{\beta}\sum_{\nu_n}e^{i\nu_n0_+}G_{\bf k}(i\nu_n+\mu)\nonumber\\
=-\int_{-\infty}^{\mu}\frac{d\nu}{\pi}\Im G_{\bf k}(\nu+i0_+),
\end{eqnarray}
rewritten at zero temperature. Then, taking into account only the first term in Eq.~(\ref{Sigma_skeleton}) and noticing that $\frac{1}{L^d}\sum_{\bf q}\Im G_{\bf q}(\nu+i0_+)=-\pi D(\nu)$ is, up to numerical prefactor, the {\it exact} density of states, we have in the large-$|{\bf k}|$ limit
\begin{eqnarray}\label{N_k_1}
\overline{N}_{\bf k}\approx\int_{0}^{\mu}\frac{d\nu}{\pi}\Im \frac{1}{\varepsilon_{\bf k}-\pi ih^2 D(\nu)}=\frac{h^2n}{\varepsilon^2_{\bf k}}+\dots,
\end{eqnarray}
where $n=\int_{0}^{\mu}d\nu D(\nu)$ by definition, and contribution to integral over $\nu$ from the negative semi-axis is strongly suppressed at large $|{\bf k}|$ because the integrand is proportional to the $\delta$-function there. Note that the first disorder-induced contact parameter coincides with $\mathcal{C}^{(1)}_h$ in (\ref{C_h}). This analysis suggests that one should keep track of the imaginary part of the self-energy while being interested in the large-momentum tail of particle distribution. Thus, we can go further considering the asymptotic behavior of the second term in Eq.~(\ref{Sigma_skeleton})
\begin{align}
h^4\int_{{\bf q},{\bf q}'}\Im\left\{G_{\bf q}(i\nu_n)G_{{\bf q}'}(i\nu_n)G_{{\bf q}+{\bf q}'-{\bf k}}(i\nu_n)\right\}_{i\nu_n\to \nu+i0_+}\nonumber\\
=3 h^4\int_{{\bf q}}\Im G_{\bf q}(\nu+i0_+)\int_{{\bf q}'}\Re \left\{G_{{\bf q}'}(\nu+i0_+)\right.\nonumber\\
\left.\times G_{{\bf q}'+{\bf k}}(\nu+i0_+)\right\}+\dots,
\end{align}
where $\int_{{\bf q}}=\int \frac{d{\bf q}}{(2\pi)^d}$. Again the first integral in the product gives $-\pi D(\nu)$, and the second one has to be calculated in the $|{\bf k}|\to \infty$ limit [where the integrand can be replaced by $1/(\varepsilon_{{\bf q}'}\varepsilon_{{\bf q}'+{\bf k}})$]
\begin{align}\label{}
\lim_{|{\bf k}|\to \infty}\int_{{\bf q}'}\Re \left\{G_{{\bf q}'}(\nu+i0_+)G_{{\bf q}'+{\bf k}}(\nu+i0_+)\right\}\nonumber\\
=c^{(2)}_d\frac{(2m)^2}{|{\bf k}|^{4-d}}+\dots.
\end{align}
Substituting the asymptotic form of the imaginary part of the second term of self-energy into Eq.~(\ref{N_k}) and repeating calculations similar to (\ref{N_k_1}), one obtains
$3c^{(2)}_d\frac{(2mh)^4n}{|{\bf k}|^{8-d}}$ [compare to the second term in (\ref{N_k_sigma})] for the subleading term in $\overline{N}_{\bf k}$.  

Apart from particle momentum distribution, the disorder affects crucially the two-point correlation function of density fluctuations at zero Matsubara frequency (zero-frequency dynamic structure factor). Large-momentum behavior of this correlator is demonstrated to be universal. Moreover, we showed that the energy of the disordered system is a linear functional of both particle momentum distribution and a zero-frequency pair correlator of density fluctuations. The latter explicitly enters the energy-pressure relation. All these conclusions remain valid independently of the fermionic system's phase, being normal Fermi liquid or superfluid, at least, in the limit of weak disorder. Even in 2$d$, where disorder of any magnitude leads to the localization transition, the pairing two-body interaction provides fermionic superfluidity. One can expect to probe our predictions in the Monte Carlo simulations taking some smooth short-ranged functions as the Gaussian average of random potential. Then, in the limit of the vanishing range, the universal behavior should be observed.

\begin{center}
	{\bf Acknowledgements}
\end{center}
We thank Profs. Yu. Holovatch and M. Dudka, for an informative introduction to different types of disorder. This work was partly supported by Project No.~0122U001514 from the Ministry of Education and Science of Ukraine.

\end{document}